\documentclass[12pt]{article}
\usepackage{psfig}
\def\be{\begin{equation}}
\def\ee{\end{equation}}
\def\beq{\begin{equation}}
\def\eeq{\end{equation}}
\def\beqar{\begin{eqnarray}}
\def\eeqar{\end{eqnarray}}
\def\barr{\begin{array}}
\def\earr{\end{array}}

\def\and{\qquad {\rm and } \qquad}

\def\slp{p \hspace{-1ex}/}
\def\slk{$k$ \hspace{-1ex}/}

\def\slk{k \hspace{-1ex}/}

\def\slp{p \hspace{-1ex}/}
\def\sls{s \hspace{-1ex}/}

\def\ttbar{$t \overline{t}~$}

\def\sp{\vec{s}_+}
\def\sm{\vec{s}_-}
\def\hp{h_+}
\def\hm{h_-}
\def\Kv{\vec{K}}
\def\pv{\vec{p}}
\def\g5{\gamma_5}
\newcommand{\nc}{\newcommand}
\nc{\jc}{\frac{1}{4}}  \nc{\sll}{S_{LL}}     \nc{\slr}{S_{LR}}
\nc{\srl}{S_{RL}}      \nc{\srr}{S_{RR}}     \nc{\vll}{V_{LL}}
\nc{\vlr}{V_{LR}}      \nc{\vrl}{V_{RL}}     \nc{\vrr}{V_{RR}}
\nc{\tll}{T_{LL}}      \nc{\tlrs}{T_{LR}}    \nc{\trl}{T_{RL}}
\nc{\trr}{T_{RR}}      \nc{\slld}{S_{LL}^D}  \nc{\slrd}{S_{LR}^D}
\nc{\srld}{S_{RL}^D}   \nc{\srrd}{S_{RR}^D}  \nc{\vlld}{V_{LL}^D}
\nc{\vlrd}{V_{LR}^D}   \nc{\vrld}{V_{RL}^D}  \nc{\vrrd}{V_{RR}^D}
\nc{\tlld}{T_{LL}^D}   \nc{\tlrd}{T_{LR}^D}  \nc{\trld}{T_{RL}^D}
\nc{\trrd}{T_{RR}^D}   \nc{\aqde}{\alpha_{qde}}
\nc{\alq}{\alpha_{\ell q}}        \nc{\alqp}{\alpha_{\ell q'}}
\nc{\alqt}{\alpha_{\ell q}^{(3)}} \nc{\alqtc}{\alpha_{\ell
q}^{(3)*}} \nc{\alqj}{\alpha_{\ell q}^{(1)}}
\nc{\alqjc}{\alpha_{\ell q}^{(1)*}} \nc{\aeu}{\alpha_{eu}}
\nc{\alu}{\alpha_{\ell u}} \nc{\aqe}{\alpha_{qe}}
\nc{\ber}{\begin{eqnarray*}} \nc{\enr}{\end{eqnarray*}}
\nc{\jmpb}{(1-\beta)/(1+\beta)} \nc{\wspR}{r}      \nc{\varx}{x}
\nc{\bt}{\beta}

\nc{\non}{\nonumber} \nc{\lspace}{\;\;\;\;\;\;\;\;\;\;}
\nc{\llspace}{\lspace \lspace}
\nc{\jnl}{\frac{1}{{\mit\Lambda}^2}} \nc{\jd}{\frac{1}{2}}
\nc{\comment}[1]{}

\begin{document}
\vskip .3cm

\begin{center}{\Large \bf \boldmath
Space-time structure of new physics with polarized beams at the
linear collider}
\vskip 1cm
{B. Ananthanarayan$^a$, Saurabh D. Rindani$^{b}$} 
\vskip .5cm

{\it $^a$Centre for High Energy Physics, 
Indian Institute of Science\\ Bangalore
560 012, India\\~ \\
$^b$Theory Group, Physical Research Laboratory\\ 
Navrangpura, Ahmedabad 380 009,
India}
\end{center}

\begin{quote}
\begin{abstract}
We approach the issue of the discovery of new physics at
high energies associated with the proposed International Linear
Collider in the presence of longitudinal as well as transverse
electron and positron beam polarization.  We determine the beam
polarization dependence and the angular
distribution of a particle of arbitrary spin in a one-particle inclusive 
final state produced in $e^+e^-$ collisions through the interference of
$\gamma$ or $Z$ amplitude with the amplitude from new interactions
having arbitrary space-time structure. We thus extend the results
of Dass and Ross proposed at the time of the discovery of neutral
currents, to beyond the standard model currents.  We also
extend the case of $e^+ e^-$ annihilation in the $s$-channel
to the production of bosons due to $t$- and $u$-channel processes.
Our work provides an approach to  model-independent determination
of the space-time structure of beyond the standard model interactions.
We briefly discuss applications of the framework to popular extensions of the
standard model, and
demonstrate that our framework is general enough to account for
certain results in the minimal supersymmetric
standard model.
\end{abstract}
\end{quote}

\section{Introduction}
The possibility of the International Linear Collider (ILC) that will
collide electrons and positrons at a large centre of mass energy,
of $\sqrt{s}=500$ GeV, or even 800 GeV is rapidly becoming a
reality~\cite{LC_SOU}.  It is also likely that the
beams can be significantly polarized and there has been a great
international effort at exploring the physics possibilities with
such a facility~\cite{POWER}.  It has been recently shown that
the availability of transverse polarization could open up the
possibility of observing CP violation in \ttbar production
due to the possibility of beyond the standard model (BSM)
interactions due to scalar and tensor type interactions~\cite{BASDR_TTBAR}.
On the other hand, it has been shown that the availability of
longitudinal polarization can significantly improve the sensitivity
to CP-violating dipole moments in this process~\cite{SDR_TOP}, and
analogously for $\tau^+\tau^-$ production~\cite{ARS} at linear
collider energies, following ideas proposed earlier in the context
of the tau-charm factories~\cite{BASDR_TAU}.   In the recent past,
we have also pointed out that BSM interactions
could lead to CP-violating anomalous triple-gauge boson 
vertices, the sensitivity to which could be enhanced by
the availability of longitudinal beam polarization~\cite{DC_SDR}
and transverse polarization~\cite{ARSB}.  It has been further
pointed out that the most general contact interactions could
also be explored exhaustively by the availability of each
type of polarization~\cite{BASDR_CONTACT1,BASDR_CONTACT2}.   
In the specific context of transversely polarized beams, some works of
interest are \cite{Hikasa,RIZZO}
These investigations
have revealed that it is fruitful to investigate in as great
generality as possible the physics that can be explored at the
ILC with polarization.

An analogous situation arose in the in the 1970's at the time the
neutral currents had just been discovered.  Dass and Ross~\cite{DR1,DR2}
considered the possibility of establishing the space-time structure
of the neutral currents by considering the interference of the
well-known QED process for electron-positron annihilation with
the then `new' physics currents to leading order.
Allowing for all possible transformation
properties of the neutral current under the Lorentz transformation,
it was possible to ask for
how the signatures would differ in the correlations
amongst the momenta of the electron, positron and one of the
reaction products, and the spins of the electron and positron,
allowing for all possible types of polarization of the initial
state particles.  The result was a table for all the possible
correlations, along with their C and P properties, which
provides a standard reference for establishing what kind of
interaction, {\it viz.} scalar, pseudoscalar, vector, axial vector
and tensor could possibly interfere with the QED amplitude and
produce possibly CP-violating signals.  

Today, one could use
the effective current induced by the BSM interaction and read off
the correlation that would result from its interference with the
QED amplitude.  However,  this would be incomplete as the standard
model (SM) amplitude results from the sum of the QED as well as
the neutral current amplitude due to the exchange of a virtual
$Z$, whose contribution is comparable to the QED amplitude at
LC energies.  It may be noted that 
the situation with the neutral current is more 
complicated than the QED case as the $Z$ has a vector as well
as an axial-vector coupling to the electron.  
It should be pointed out here that our results are
rather general; as we will show, angular correlations
involving sparticles in the minimal supersymmetric standard
model (MSSM) can also be accounted for in this framework.

Indeed, in the case of \ttbar it was possible
to guess that there would be no CP-violating observable with
only transversely polarized beams if all BSM interactions were
of the vector or axial-vector type, which was borne out by
explicit computations, with amplitudes for BSM physics given
in an effective Lagrangian framework \cite{BASDR_TTBAR}, even 
though a general
discussion of the Dass and Ross type was not available when
both QED and $Z$ exchange contributions are included.  

In this work, we will
now provide such a general discussion and provide an explicit
table for the correlations of interest in the presence of
both vector and axial-vector couplings of the intermediate
vector boson.  This has the additional advantage of not
having to appeal to any effective low-energy theory or
an effective Lagrangian framework.  We will, however, 
point out the connections of the present work to those
presented in ref.~\cite{BASDR_TTBAR}.  The framework
employed therein was the one presented in ref.~\cite{Grzadkowski},
which is a restrictive one, where a special class of couplings
is considered.

Since we restrict ourselves to the the measurement of the
energy-momentum of a single particle in the final state, we exclude the
possibility that the spin of the final state particle is measured. 
The question we are asking is how far can we go in the analysis of the new
physics using only the energy and momentum of a single final-state
particle, but equipped with polarized $e^-$ and $e^+$ beams. It should
be noted that we implicitly include the case when the single final-state 
particle studied could be the decay product of an unstable particle.

Considerations of this type cannot applied directly
to the situation when the SM contributions
to a process come from $t$- and $u$-channel contributions. However, use can
be made of our results with appropriate modifications. We
provide a discussion on this subject.  

In Sec. 2 we describe the calculation of single-particle 
angular distributions
arising from the interference of $s$-channel exchange SM terms with the
BSM terms of arbitrary space-time structure.
In Sec. 3, we discuss to what extent form factors
introduced in Sec. 2 may be determined from the angular
distributions. Sec. 4 is devoted to a discussion of the CP properties
of the various correlations. A discussion of SM processes with $t$- and
$u$-channel exchanges is given in Sec. 5. 
Some applications of our formalism are presented in Sec. 6.
followed by our conclusions in Sec. 7.

\section{Computation of correlations}

We consider the one-particle inclusive process
\begin{equation}\label{process}
e^-(p_-) + e^+(p_+) \to H(p) + X,
\end{equation}
where $H$ is a final state particle, whose momentum $p$ is measured, but not
the spin, and $X$ is an inclusive state. 
The process is assumed to occur through an $s$-channel
exchange of a photon and a $Z$ in the SM, and through a  new current whose 
coupling to $e^+e^-$ can be of the type $V,A$, or $S,P$, or $T$.  
	
Since we will deal with a general case without specifying the nature or
couplings of $H$, we do not attempt to write the amplitude for the process
(\ref{process}). We will only obtain the general form, in each case of
the new coupling, of the contribution to the angular distribution of $H$
from the interference of the SM amplitude with the new physics
amplitude.  

Following Dass and Ross \cite{DR1,DR2}, we calculate the relevant factor in
the interference between the standard model currents with the 
BSM currents as
\begin{equation}\label{trace}
{\rm Tr}[(1-\g5 \hp + \g5 \sls_+)\slp_+\gamma_\mu(g_V^e-g_A^e \gamma_5)
(1+\g5 h_-+\g5 \sls_-)\slp_-\Gamma_i]H^{i\mu }.
\end{equation}
Here $g_V^e, g_A^e$ are the vector and axial-vector couplings of the
photon or $Z$ to the electron current, and $\Gamma_i$ is the
corresponding coupling to the new physics current, $p_{\pm}$ are 
the four-momenta of $e^{\pm}$, $h_{\pm}$ are the
helicities (in units of $\frac{1}{2}$) 
of $e^{\pm}$, and $s_{\pm}$ are respectively their transverse polarizations. 
For ease of comparison, we have sought to stay with the notation 
for refs.~\cite{DR1,DR2}, with some exceptions which we
spell out when necessary.  We should of course add the
contributions coming from photon exchange and $Z$ exchange, with the
appropriate propagator factors. However, we give here the results for
$Z$ exchange, from which the case of photon can be deduced as a special
case. The tensor $H^{i\mu }$ stands for the interference between the
couplings of the final state to the SM current and the new physics
current, summed over final-state polarizations, and over the phase space
of the unobserved particles $X$. It is only a function of the the
momenta $q=p_-+p_+$ and $p$. The implied summation over $i$
corresponds to a sum over the forms $V, A, S, P, T$, together with any
Lorentz indices that these may entail. 
We will continue to use for the tensor $H^{i\mu }$ 
the term ``hadronic tensor", which is of historical origin,
and whose use was relevant to the case 
when the final states considered were hadronic.

We now determine the forms of the matrices $\Gamma_i$ and the 
tensors $H^{i\mu }$ in the various
cases, using only Lorentz covariance properties. 
Our additional currents are as in refs.~\cite{DR1,DR2},  
except for the sign of $g_A$ in the following.  We explicitly
note that in 
our convention is $\epsilon_{0123}=+1$.  Our results for
the $\epsilon$ terms differ from those in refs.~\cite{DR1,DR2} by a sign,
which we attribute to a different sign convention for the
$\epsilon$ symbol, a convention that is not explicitly spelt out in 
\cite{DR1,DR2}. We set the electron mass to zero.  Consider now
the three cases:

\medskip

\noindent\underline {1. Scalar and Pseudoscalar case}: 

In this case, there is
no free Lorentz index for the leptonic coupling. 
Consequently, we can write it as 
\begin{equation}
\Gamma = g_S + i g_P \gamma_5.
\end{equation}
The tensor $H^{i\mu }$ for this case has only one index, viz., $\mu$.
Hence the most general form for $H$ is
\begin{equation}
H^{S}_\mu  = F(q^2,p\cdot q) p_\mu,
\end{equation}
where F is a function of the Lorentz-invariant quantities $q^2$ and
$p\cdot q$.

\medskip

\noindent\underline {2. Vector and Axial-Vector case}: 

The leptonic
coupling for this case can be written as
\begin{equation}
\Gamma_\mu = \gamma_\mu (g_V -  g_A \gamma_5).
\end{equation}
Note that we differ from Dass and Ross \cite{DR1,DR2} in the sign of the
$g_A$ term.
The tensor $H$ for this case has two indices, and can be written as
\begin{equation}
H^V_{\mu\nu} =  -g_{\mu\nu} W_1(q^2,p\cdot q) + p_\mu p_\nu
W_2(q^2,p\cdot q) + \epsilon_{\mu\nu\alpha\beta}q^{\alpha}p^\beta
W_3(q^2,p\cdot q),
\end{equation}
where now there are three invariant functions, $W_1, W_2, W_3$.

\medskip

\noindent\underline {3. Tensor case}: 

In the tensor case, 
the leptonic coupling is 
\begin{equation}
\Gamma_{\mu\nu} = g_T \sigma_{\mu\nu}.
\end{equation}
The tensor $H$ for this case can be written in terms of the four
invariant functions $F_1, F_2, PF_1, PF_2$ as
\begin{equation}
\begin{array}{lcl}
H^T_{\mu\rho\tau}& = & (q_\rho  p_\tau - q_\tau p_\rho ) p_\mu
F_1(q^2,p\cdot q) + ( g_{\rho\mu} p_\tau - g_{\tau\mu} p_\rho )
F_2(q^2,p\cdot q)\\&& + \epsilon_{\rho\tau\alpha\beta} p^\alpha q^\beta p_\mu
PF_1(q^2,p\cdot q) + \epsilon_{\rho\tau\mu\alpha} p^\alpha
PF_2(q^2,p\cdot q).
\end{array}
\end{equation}

\medskip

We next substitute the leptonic vertices $\Gamma$ and the respective
tensors $H^i$ in  (\ref{trace}), and
evaluate the trace in each case. We present the results in Tables
1-6, with
$\vec K\equiv (\vec{p}_- - \vec{p}_+)/2= E \hat{z}$, 
where $\hat{z}$ is a unit vector
in the z-direction, $E$ is the beam energy,
and $\vec{s}_\pm$ lie in the x-y plane.

In Tables 1, 3 and 5 the results presented in ref.~\cite{DR1,DR2} 
are reproduced in our convention, since these are for the case 
of $g_V^e$ alone, 
which was the case considered by these authors for the interference
of QED amplitudes with physics due to the then undetermined amplitude of the
neutral current due to $Z$. 
The corresponding tables for $g_A^e$ are now
presented in Tables 2, 4 and 6, for cases of scalar-pseudoscalar,
vector-axial-vector and tensor couplings respectively.  

In Tables 1-6 are also given the charge conjugation C and parity P
properties of the various correlations, under the assumption that the
final-state particle observed is self-conjugate, viz., $H=\overline H$. 
If it is not
self-conjugate, then the C factor given in the tables would apply to the
sum of the cross sections for production of $H$ and $\overline H$.
The difference of these cross sections would take a C factor of the
opposite sign.

\begin{table}\label{ps_gVe_table}
\begin{center}
\begin{tabular}{||c|c|c|c||}\hline
Term & Correlation & ${\rm P}$ & ${\rm C}$ \\ \hline \hline
${\rm Im}\, (g_P F)$ & $-2 E^2 \,(\sp-\sm)\cdot \pv$ & $-$ & $-$ \\ 
${\rm Im}\, (g_S F)$ & $2 E \, [\Kv \cdot  (\sp+\sm)\times \pv]$ & $+$ & $-$ \\
${\rm Re}\, (g_S F)$ & $2 E^2 \, \pv \cdot (\hp \sm - \hm \sp)$ & $+$ & $-$ \\
${\rm Re}\, (g_P F)$ & $-2 E \, [\Kv \cdot  (\hp \sm + \hm \sp] \times \pv)$ & 
						$-$ & $-$ \\ \hline
\end{tabular}
\caption{List of $S,P$ correlations for $g_{V}^e$}
\end{center}
\end{table}

\begin{table}\label{ps_gAe_table}
\begin{center}
\begin{tabular}{||c|c|c|c||}\hline
Term & Correlation & ${\rm P}$ & ${\rm C}$ \\ \hline \hline
${\rm Im}\, (g_P F)$ & $2 E^2 \,(\hp \sm+\hm \sp)\cdot \pv$ & $+$ & $+$ \\ 
${\rm Im}\, (g_S F)$ & $2 E \, [\Kv \cdot  (\hp\sm-\hm\sp)\times \pv]$ 
							& $-$ & $+$ \\
${\rm Re}\, (g_S F)$ & $2 E^2 \, \pv \cdot (\sp +  \sm)$ & $-$ & $+$ \\
${\rm Re}\, (g_P F)$ & $2 E \, [\Kv \cdot  (\sp -  \sm) \times \pv]$ & $+$ &
                            				$+$ \\ \hline
\end{tabular}
\caption{List of $S,P$ correlations for $g_{A}^e$}
\end{center}
\end{table}

\section{Determination of form factors from correlations}

For simplicity and for the purposes of this section, we do not 
consider simultaneous presence of both
longitudinal and transverse polarizations. Thus, we will discuss the
possibilities of unpolarized beams, pure longitudinal polarization and
pure transverse polarization. We would like to investigate how many of
the independent form factors can be determined by the independent
angular correlations available.

By examining the
tables it can be seen that in the absence of polarization, the only
correlations which survive are in the case of $V, A$ interactions.
Even in that case, it is only possible to get information on the
quantities Re$(g_V W_1)$,  Re$(g_V W_2)$ and Im$(g_A W_3)$ in association 
with $g_V^e$, and Re$(g_A W_1)$,  Re$(g_A W_2)$ and Im$(g_V W_3)$ in
association with $g_A^e$. However, they give only three independent
angular distributions, and as such, only 3 combinations of these 6
quantities can be determined in the absence of polarization.

In the presence of longitudinal polarization alone, it is possible to
get information on 3 more combinations
of form factors, proportional to $(h_+ -  h_-)$, using either
$e^-$ or $e^+$ polarization in case of $V$ and $A$ couplings. It is not
necessary to have both $e^-$ and $e^+$ beams polarized. 

It is clear why only $V$ and $A$ couplings can contribute in the absence
of polarization, or in the presence of only longitudinal polarization.
We are concentrating on interference between the $V,\, A$ SM 
contribution\footnote{Our framework being sufficiently
general may readily be generalized to $V,\, A$ contributions
in other theories, e.g., the MSSM.  For a discussion of
certain results therein, see Sec. 5, 6.}
and the new physics contribution. Hence the terms which survive should
correspond to the same combination of $e^-$ and $e^+$ helicities. Since
$V,\, A$ amplitudes have a different helicity structure as compared
to the amplitudes coming from the $S, P, T$ structure from 
BSM, they do not interfere.

\begin{table}\label{VA_gVe_table}
\begin{center}
\begin{tabular}{||c|c|c|c||}\hline
Term & Correlation& ${\rm P}$ & ${\rm C}$ \\ \hline \hline
${\rm Re} \,( g_V W_1)$ & $-4 E^2 (\hp \hm -1) $ & $+$ & $+$ \\
${\rm Re} \,( g_A W_1)$ & $4 E^2 (\hp - \hm ) $ & $-$ & $-$ \\
${\rm Re} \,( g_V W_2)$ & $-2[2 E^2\, \pv\cdot \sm \pv \cdot \sp +
       (\Kv\cdot\Kv
   \, \pv\cdot\pv -(\pv\cdot \Kv)^2)(\hp\hm-1-\sp\cdot \sm)]$ & $+$ & $+$ \\
${\rm Re} \, (g_A W_2)$ & $ 2 (\Kv\cdot\Kv\, \pv\cdot\pv -(\pv\cdot \Kv)^2) (\hp-\hm)$
& $-$ & $-$ \\
${\rm Im}\, (g_V W_3)$ & $ 8 E^2 (\pv\cdot\Kv)  (\hp-\hm)$ & $-$ & $+$ \\
${\rm Im}\, (g_A W_3)$ & $-8 E^2 (\pv \cdot \Kv)  (\hp \hm-1)$ & $+$ & $-$ \\
${\rm Im}\, (g_A W_2)$ & $2E(\pv\cdot \sp [\Kv\cdot \sm\times \pv] +
                        \pv\cdot \sm [\Kv\cdot \sp\times \pv] )$ & $-$ & $-$ \\
\hline
\end{tabular}
\caption{List of $V,A$ correlations for $g_{V}^e$}
\end{center}
\end{table}

\begin{table}\label{VA_gAe_table}
\begin{center}
\begin{tabular}{||c|c|c|c||}\hline
Term & Correlation & ${\rm P}$ & ${\rm C}$ \\ \hline \hline
${\rm Re}\, (g_V W_1)$ & $4 E^2(\hp - \hm ) $ & $-$ & $-$ \\
${\rm Re}\, (g_A W_1)$ & $-4 E^2(\hp \hm -1) $ & $+$ & $+$ \\
${\rm Re} \, (g_V W_2)$ & $ 2 (\Kv\cdot\Kv
         \, \pv\cdot\pv -(\pv\cdot \Kv)^2) (\hp-\hm)$
& $-$ & $-$ \\
${\rm Re} \, (g_A W_2)$ & $-2[-2 E^2 \pv\cdot \sm \pv \cdot \sp +
       (\Kv\cdot\Kv\, \pv\cdot\pv -
	(\pv\cdot\Kv)^2)(\hp\hm-1+\sp\cdot \sm)]$ & $+$ & $+$ \\
${\rm Im}\, (g_V W_3)$ & $-8 E^2(\pv \cdot \Kv)  (\hp \hm-1)$ & $+$ & $-$ \\
${\rm Im}\, (g_A W_3)$ & $ 8 E^2(\pv\cdot\Kv)  (\hp-\hm)$ & $-$ & $+$ \\
${\rm Im}\, (g_V W_2)$ & $-2E(\pv\cdot \sp [\Kv\cdot \sm\times \pv] +
                        \pv\cdot \sm [\Kv\cdot \sp\times \pv] )$ & $-$ & $-$ \\
								\hline
\end{tabular}
\caption{List of $V,A$ correlations for $g_{A}^e$}
\end{center}
\end{table}

The case with transverse polarization is somewhat different. Since
transverse polarization corresponds to a linear combination of two
different helicities, it is possible for interference terms of different
helicity structures to survive.

With only transverse polarization, it can be seen from the tables that 
one can get information on all 4 combinations of $S,P$ type couplings, 3
combinations of $V,A$ type couplings involving only $W_2$, 
and 8 combinations of $T$ type
couplings. Of course, if $S,P$ and $T$ couplings are present
simultaneously, then only an overall total of 8 combinations can be
determined. 

In the case of $S,P$ and $T$ couplings, it is sufficient to have either
$e^-$ or $e^+$ beams polarized. In case of $V,A$, both beams have to be
polarized, or the effect vanishes.
It is interesting to note that all the correlations in the latter case are
symmetric under the interchange of $\vec s_+$ and $\vec s_-$.

\begin{table}\label{T_gVe_table}
\begin{center}
\begin{tabular}{||c|c|c|c||}\hline
Term & Correlation & ${\rm P}$ & ${\rm C}$ \\ \hline \hline
${\rm Im}\, (g_T F_1)$ & $-8 E^2 \pv \cdot \Kv [\pv\cdot(\hp\sm-\hm\sp)]$ & 
					$+$ & $+$ \\
${\rm Im}\,(g_T  F_2)$ & $4 E^2 \pv\cdot(\hp\sm+\hm\sp)$ & $+$ & $+$ \\
${\rm Im}\,(g_T  PF_1)$ & $8 E \pv\cdot \Kv [\Kv\cdot(\hp\sm+\hm\sp)\times \pv]$ &
							        $-$ & $+$ \\
${\rm Im}\, (g_T PF_2)$ & $4E  [\Kv\cdot(\hp\sm-\hm\sp)\times \pv]$ &
							        $-$ & $+$ \\
${\rm Re}\, (g_T F_1)$ & $8 E \pv\cdot \Kv [\Kv\cdot(\sp+\sm)\times \pv]$ &
							        $+$ & $+$ \\
${\rm Re}\, (g_T F_2)$ & $ 4E  [\Kv\cdot(\sp-\sm)\times \pv]$ &
							        $+$ & $+$ \\
${\rm Re}\, (g_T PF_1)$ & $-8E^2 \pv\cdot\Kv[\pv\cdot(\sp-\sm)]$ & $-$ & $+$ \\
${\rm Re}\, (g_T PF_2)$ & $4 E^2 \pv\cdot(\sp+\sm)$ & $-$ & $+$ \\ \hline
\end{tabular}
\caption{List of $T$ correlations for $g_{V}^e$}
\end{center}
\end{table}

\begin{table}\label{T_gAe_table}
\begin{center}
\begin{tabular}{||c|c|c|c||}\hline
Term & Correlation & ${\rm P}$ & ${\rm C}$ \\ \hline \hline
${\rm Im}\, (g_T F_1)$ & $-8E^2 \pv\cdot\Kv[\pv\cdot(\sp+\sm)]$ & $-$ & $-$ \\
${\rm Im}\, (g_T F_2)$ & $-4 E^2 \pv\cdot(\sp-\sm)$ & $-$ & $-$ \\
${\rm Im}\, (g_T PF_1)$ & $-8 E \pv\cdot \Kv [\Kv\cdot(\sp-\sm)\times \pv]$ &
							        $+$ & $-$ \\
${\rm Im}\, (g_T PF_2)$ & $4 E [\Kv\cdot(\sp+\sm)\times \pv]$ &
							        $+$ & $-$ \\
${\rm Re}\,  (g_T F_1)$ & $8 E \pv\cdot \Kv [ \Kv\cdot(\hp\sm-\hm\sp)\times \pv]$ &
							        $-$ & $-$ \\
${\rm Re}\,  (g_T F_2)$ & $-4 E [\Kv\cdot (\hp\sm+\hm\sp)\times \pv]$ &
							        $-$ & $-$ \\
${\rm Re}\, (g_T PF_1)$ & $8 E^2 \pv \cdot \Kv [\pv\cdot(\hp\sm+\hm\sp)]$ & $+$ & $-$ 
								\\
${\rm Re}\, (g_T PF_2)$ & $4 E^2 \pv\cdot(\hp\sm-\hm\sp)$ & $+$ & $-$ \\
								\hline
\end{tabular}
\caption{List of $T$ correlations for $g_{A}^e$}
\end{center}
\end{table}

An inspection of the momentum and spin correlations of Tables
5 and 6 reveals that apart from
overall multiplicative factors including those of energy and
scalar products of momenta, the momentum and spin correlations
are precisely those that are found in Tables 1 and 2.                  

The counting of the number 
of independent correlations for the vector and axial-vector cases turns
out to be subtle.  First consider having only non-zero $g_{V}^e$.
The number of independent form factors that contribute to the
correlations is 7, and the number of correlations
is 7. When we 
consider the case when only $g_A^e$ is non-zero,  there are again
7 independent form factors, of which 6 are common to the previous
list. More explicitly, ${\rm Im}\, (g_A W_2)$ contributes to
the correlations only via $g_V^e$, while  
${\rm Im}\, (g_V W_2)$ contributes to
the correlations only via $g_A^e$, noting however that
the resulting correlation is exactly the same apart from the sign. 
Indeed, the momentum and spin correlations also
have the structure of 6 being in common and 1 each from each of
the two cases above;  
the correlation due to 
${\rm Re}\, (g_V W_2)$ via $g_V^e$ differs from the correlation
resulting via $g_A^e$.
Thus, although the independent form factors and
the corresponding momentum and spin correlations are not in
one to one correspondence, each set adds up separately
to 8. The total number of independent form factors is 12, and even if
both longitudinal and transverse polarizations are used, 4 of these,
viz., Im$(g_VW_1)$, Im$(g_AW_1)$, Re$(g_VW_3)$ and Re$(g_AW_3)$ do not
enter the distributions, and remain undetermined.

\section {CP properties of correlations}

We might like to use the behaviour of the differential cross
section to construct asymmetries which can test symmetry properties like
CP. Tables 1-6 may be employed to make some predictions for what to
expect.

There are some general deductions we can make in the special case when
the final-state is a two particle state. Within that, we consider two
possibilities:

\medskip
\noindent \underline {Case 1: $H=\overline H$}
The simplest case to consider is when 
$H$ is self-conjugate, i.e., $H=\overline H$. 

In the absence of polarization, 
of the 3 independent angular distributions, Tables 3 and 4 show that
only one, viz.,  $\pv \cdot \Kv = E \vert \pv \vert \cos\theta $,
where $\theta$ is the angle made by $\pv$ with the $e^-$ beam direction,
is CP violating. It accompanies ${\rm Im}(g_V W_3)$ or ${\rm Im}(g_A
W_3)$, and needs the absorptive part of some amplitude to be nonzero,
consistent with the CPT theorem.

With longitudinal polarization alone, of the 
three additional combinations possible, one is CP violating, again proportional
to $\pv \cdot \Kv = E \vert \pv \vert \cos\theta $. 

It may be deduced from the tables that for the case of only transverse
polarization, CP-violating correlations arise
in the case of the terms $g_V^e{\rm Im}(g_SF)$, $g_A^e{\rm Re}(g_SF)$,
$g_V^e{\rm Re}(g_TPF_1)$, $g_V^e{\rm Re}(g_TPF_2)$, $g_A^e{\rm
Im}(g_TPF_1)$ and $g_A^e{\rm Im}(g_TPF_2)$.
As in
the case of QED, even in the electro-weak theory, the interference
of new currents cannot lead to a CP-violating correlation due to
vector and axial-vector type couplings. 
With transverse polarization, one can
probe CP violation only due to new physics of the $S,\, P$ and $T$
type.

\medskip

\noindent \underline{Case 2: $H \neq \overline H$}

As mentioned earlier, in this case, the C properties in the tables refer
to the sum 
\begin{equation}
\Delta \sigma^+ = \Delta\sigma + \Delta\bar \sigma,
\end{equation}
where $\Delta\sigma$ and $\Delta\bar \sigma$ are partial cross sections 
corresponding respectively to $H$ and $\overline H$
production. The difference of these, 
\begin{equation}
\Delta \sigma^- =  \Delta\sigma - \Delta\bar \sigma,
\end{equation}
will have the opposite C property.

We now consider two special cases: when the final state consists of
a pair of conjugate particles $H\overline H$, and when it consists of two
particles $H\overline H'$, where $H' \neq H$.

\medskip

\noindent \underline{Case 2a: $X\equiv \overline H$}

In this 
case, $\Delta\bar \sigma$ is obtained from $\Delta\sigma$ by simply the
reversal of the sign of $\vec p$. In this case,
the correlations which are odd in $\vec p$ vanish in $\Delta \sigma^+$.
They survive in $\Delta \sigma^-$, but have the opposite C property as
compared to the one shown in the corresponding entry in the relevant
table. The correlations which are even in $\vec p$, on the other hand, 
 vanish in $\Delta\sigma^-$,
but survive in $\Delta \sigma^+$, and have the same C property as
the one shown in the table.

We can deduce the following for the case of the $H\overline H$ final state.
Since the only correlations shown as CP odd in Tables 3 and 4 are 
linear 
in $\vec p$, they do not survive in $\Delta \sigma^+$. Hence there are
no CP-odd correlations in the $V,A$ case. This statement is true
regardless of whether there is polarization or not. 
In the $S,P$ case, all terms are linear in $\vec p$. Hence they all
vanish in $\Delta\sigma^+$. They do survive in $\Delta\sigma^-$, and
have CP properties opposite of those shown in Tables 1,2. Thus,
considering only transverse polarizations, the terms corresponding to
$g_V^e{\rm Im}(g_PF)$ and $g_A^e{\rm Re}(g_PF)$ are the ones odd under
CP, whereas the terms corresponding to $g_V^e{\rm Im}(g_SF)$ and
$g_A^e{\rm Re}(g_SF)$ are even under CP.
Finally, in the $T$ case, 
one can check that the terms corresponding to $g_V^e{\rm Re}(g_TF_2)$
$g_V^e{\rm Re}(g_TPF_1)$, $g_A^e{\rm Im}(g_TF_2)$ and $g_A^e{\rm
Im}(g_TPF_1)$ are CP 
odd and the remaining, viz., $g_V^e{\rm Re}(g_TF_1)$
$g_V^e{\rm Re}(g_TPF_2)$, $g_A^e{\rm Im}(g_TF_1)$ and $g_A^e{\rm
Im}(g_TPF_2)$ are CP even.
It is interesting to note that in all the above, the CP-odd cases 
correspond to a combination $(\vec s_+ - \vec s_-)$ of the $e^+$ and 
$e^-$ spin vectors. The reason is that since in the cm frame the
momentum vectors of $H$ and $\overline H$ are equal and opposite, $\vec p$ is
equivalent to $(\vec p_H - \vec p_{\overline H})$, and hence even under CP, as
is $\vec K$. The
only quantity odd under CP is $(\vec s_+ - \vec s_-)$, and hence necessary
for the term to be CP odd.

We can also consider the possibility that there are no loop effects or
final-state interactions. In that case, the $g_i,\,(i=S,P,V,A,T)$ are
all real, and the form factors $F_1$ and $W_3$ are purely imaginary, the
rest of the form factors being real. 
Then, the only CP-odd contributions are $g_A^e{\rm Re}(g_PF)$ in the
$S,P$ case, and the terms corresponding to $g_V^e{\rm Re}(g_TF_2)$ 
and $g_A^e{\rm Im}(g_TPF_1)$ in the tensor case. 

\medskip

\noindent \underline{Case 2b: $X\equiv\overline H'$, $H\neq H'$}

We now consider the possibility that the final-state is a
two-particle state, of the form $H \overline H'$, where $H'$ is not the same
as $H$. In such a case, we cannot rule out a priori either the CP-even
or the CP-odd combinations.  However, 
we can make definite statements about CP-odd
terms only in the special case that we restrict ourselves to tree-level
contributions, and there are no loop contributions or final-state
interactions. The simplification in this case is that the effective
Lagrangian for all interactions can be taken to be Hermitian. As a
result, the couplings and form factors contributing to $\Delta\bar
\sigma$ would be complex conjugates of those contributing to
$\Delta\sigma$. Hence the real parts of these couplings would be
equal, and the imaginary parts equal in magnitude, and opposite in sign. 
Thus, for a term labelled as CP even in one of the tables, the
combination $\Delta\sigma^-$ would be CP odd, and would survive only
provided it comes with the imaginary part of couplings. For a term
labelled CP odd, the combination $\Delta\sigma^+$ would be CP odd, and
would survive provided the corresponding couplings came with their real
part. Thus, for example, in the $V,A$ case, the only surviving CP odd
combinations correspond to the last rows in Tables 3 and 4, and the
corresponding combination of couplings would be ${\rm Im}\left[\left(g_V^eg_A
- g_A^e g_V\right)W_2\right]$. In the $S,P$ and $T$ cases, more
possibilities survive, and we do not list them here.

\section{Processes with $t$-channel  and $u$-channel exchanges}

So far we have dealt with a scenario where the SM interactions take
place through $s$-channel $\gamma$ and $Z$ exchanges. This is most
suitable for production of particles which have not direct coupling to
$e^-$ and $e^+$. However, for production of gauge bosons in the SM, which
couple directly to $e^+e^-$, there would be a $t$-channel and/or
$u$-channel lepton exchange. 

It was also assumed that the new physics processes can
also be represented by the $s$-channel exchange of a new particle, or by
contact interactions. 

The considerations of the early treatment could be carried over to these
cases if one rewrites  $t$- and
$u$-channel interactions or amplitudes as effective interactions, where
the corresponding couplings, or rather form factors,  depend on $t$ or $u$.
Even in this case we have to deal with this new situation where the
correlations obtained in the foregoing would get modified because of the
$t$ and $u$ dependence of the form factors.

We first study how  our discussion of the separate cases of $S,P$, $V,A$
and $T$ couplings can be adapted to this case.
The crucial factor in this
adaptation is the fact that for $m_e=0$, the only contributions which
survive correspond to opposite $e^-$ and $e^+$ helicities. For, any
final-state particles which may be emitted from an electron line with a
flip of electron helicity (as for example a Higgs boson) will have
vanishing coupling in the limit of $m_e=0$. We are thus left in the most
general case with only chirality-conserving combinations of Dirac
matrices, sandwiched between electron and positron spinors of opposite
helicities. Such a combination of Dirac matrices is a product of odd
number of them. For massless spinors, they can always be reduced to a
linear combination of $\gamma_{\mu}$ and $\gamma_{\mu} \gamma_5$. We are
thus back to the case of $V$ and $A$ couplings in the $s$
channel considered in the preceding, except that the coefficients $g_V^e$
and $g_A^e$ would now be replaced by something more complicated. In fact,
they could contain tensors constructed out of momenta occurring in the
process. It is possible to absorb these tensors into the ``hadronic" vertices.
The final result would be that we could still use the tables we have
obtained so far, with appropriate redefinitions of $g_V^e$, $g_A^e$ and the
form factors.  While this is a general feature, it is
a discussion in very general terms not possibl. We will, instead, 
illustrate this for a special case of the process $e^+e^- \to \gamma Z$,
which occurs in the SM through $t$- and $u$-channel exchange of an electron.

The matrix element for the process $e^-(p_-)+e^+(p_+)\to
\gamma_{\alpha}(k_1) + Z_{\beta}(k_2)$ can be written as 
\begin{equation}
M = e\bar v (p_+)\left[ \frac{1}{t} \gamma^{\beta}(g_V^e-g_A^e \gamma_5)(\slp
_- - \slk_1) \gamma^{\alpha} +
 \frac{1}{u} \gamma^{\alpha}(\slp_- - \slk_2)
\gamma^{\beta}(g_V^e-g_A^e \gamma_5) \right]u(p_-),
\end{equation}
where $t=(p_--k_1)^2$ and $u=(p_--k_2)^2$.
After some algebra, the equation above can be rewritten as 
\begin{equation}
M= e\bar v(p_+) \left[ \gamma_{\mu} (g_V^e-g_A^e\gamma_5) T_1^{\mu\alpha
\beta} + \gamma_{\mu} ( g_A^e - g_V^e \gamma_5) T_2^{\mu \alpha \beta} \right]
u(p_-),
\end{equation}
where 
\begin{equation}
T_1^{\mu\alpha\beta} = g^{\mu\beta}\left(\frac{2p_-^{\alpha}}{t}
- \frac{2p_+^{\alpha}}{u}\right) + (- g^{\mu\alpha}
k_1^{\beta} + g^{\alpha\beta} k_1^{\mu})\left(\frac{1}{t} -
\frac{1}{u}\right),
\end{equation}
and 
\begin{equation}
T_2^{\mu\alpha\beta}= -i \epsilon^{\mu\alpha\beta\lambda} k_{1\lambda}
\left( \frac{1}{t} + \frac{1}{u} \right).
\end{equation}

As can be seen from the above, we are in a position to use the formalism
of the previous sections, except that the SM contribution has a
different form for the ``hadronic" vertex, while the leptonic vertex
remains of the $V$ and $A$ form.
We could thus go ahead and consider the cross term of this with the new
interactions. So far as the leptonic tensor concerned, we only have to
modify what we use for $g_V^e$ and $g_A^e$. The hadronic tensor would now be
more complicated and would have to be computed using the cross term
of  $T_1$ and $T_2$ above with the tensors arising in the BSM
interactions. There is an apparent problem in that the ``hadronic"
tensor will involve leptonic momenta, as for example, in $T_1$. However,
on summing over final state polarizations, the leptonic momenta will be
contracted appropriately to give Lorentz scalars like $p_-.k_1$,
$p_-.k_2$, etc., which can be rewritten in terms of $s$, $t$ and $u$, and
will contribute to the $t$- and $u$-dependence of the effective 
new form factors.

The next important issue we have to deal with, then, is the $t$ and $u$
dependence of form factors, which was assumed absent in the simple
treatment which led to the Tables 1-6. If we treat processes involving
$t$- and $u$-channel exchanges, the propagators corresponding to these
exchanges will contribute to the $t$ and $u$ dependence of the form
factors.

In the original inclusive process (\ref{process}) which we consider, the
kinematic variables $t$ and $u$ may be written as
\begin{equation}\label{tu}
t =  (p_- - p)^2 = -2 p_-\cdot p + p^2, \; u =  (p_+ - p)^2 = -2
p_+\cdot p + p^2.
\end{equation}
Since $t+u =  -2 p\cdot q + 2 p^2$, it is only the dependence on the 
combination $t-u = 4
\vec p \cdot \vec K$ which is new when we allow form factors to depend
on $t$ and $u$. Thus, in such a case, the correlations of Tables 1-6
would get additional dependence on $\vec p \cdot \vec K$. However, they
would not get any additional dependence on triple product of vectors, or
on spin.
This dependence on $\vec p \cdot \vec K$ would also have important
consequences when we discuss the CP properties of the correlations.

In the case when $H=\overline H$, a CP transformation simply interchanges $t$
and $u$. Then the CP behaviour of a certain correlation is even or odd
depending on the product of the CP phase factor
obtained from the relevant table and a factor $\pm 1$ coming from the
behaviour of the amplitude under $t$-$u$ interchange.

In the example of the process $e^+e^- \to \gamma Z$ considered above,
the form factor $T_1$ is odd under the interchange of $t$ and $u$. When
the new interactions are CP violating, it
could give rise to a CP-odd correlation of one of the forms listed in
Tables 3 and 4 as being CP even. $T_1$ is  proportional to $(t-u)$, and
the consequent factor of $\vec K \cdot \vec p$, as it turns out,
multiplies the last entry in Table 4, giving rise to a CP-odd
correlation when the new interaction considered is either an anomalous
$\gamma\gamma Z$ coupling or a contact $e^+e^-\gamma Z$ interaction. We
refer to \cite{ARSB,BASDR_CONTACT1} for details.

The same CP-odd correlation arises in a different context of
supersymmetry in the process $e^+e^- \to \tilde\chi_i \tilde\chi_j$, $i\neq j$, 
where $\tilde\chi_i$ are neutralinos in the theory, which are self-conjugate.
Here the CP-odd terms arise in the cross terms between the $s$-channel
and the $t$- and $u$-channel production diagrams \cite{Choi,Bartl1}.

In the case when $H\neq \overline H$, and the final state is $H\overline H$,
a CP transformation keeps $t$ and $u$ unchanged. Hence the earlier
discussion of CP properties of correlations for this case goes through  
unchanged. 

In the case of a final state $H\overline H'$ ($H\neq H'$), one would
compare the cross section combinations $\Delta\sigma_{\pm}$, which are
linear combinations of cross sections for the production of $H\overline H'$
and $\overline H H'$. In this case, under CP, $t$ and $u$ for the first 
process would go respectively to $t$ and $u$ for the second process.
Thus, the discussion of CP properties given earlier for this case goes
through. The specific example of chargino pair production in
supersymmetry would serve as an illustration for this case, which is
discussed in the next section.

\section{Some applications}

In our earlier work on $e^+ e^-\to t\overline{t}$~\cite{BASDR_TTBAR}
we had considered the possibility of CP violation arising from
four-Fermi interactions due to an effective Lagrangian framework with
transversely polarized beams. We found
that there was no CP-violating observable possible with $V$ and $A$ type
of four-Fermi interactions. However, CP-violating observables did exist
for $S,P$ and $T$ interactions. The negative result for $V,A$ interactions
is clear from our discussion of the case 2a of Sec. 4.
For the $S,P$ and $T$ case, it is possible to have CP-odd observables,
and of the possible ones listed therein, the ones which occur
in the special case of lowest-dimensional observables are Re$(g_P F)$
and Re$(g_T F_2)$, corresponding respectively to the four-Fermi couplings 
Im$(S_{RR})$ and Im$(T_{RR})$ in the notation of \cite{BASDR_TTBAR}. The
special features observed in that work, viz., that the four-Fermi 
scalar coupling
terms occur with only the $g_A^e$ coupling at the electron vertex, and
that the tensor coupling terms occur with only the $g_V^e$ at the
electron vertex, are borne out by our general results. 

Rizzo~\cite{RIZZO} has considered probing extra-dimensional models
using transverse polarization. The key observation is that an
azimuthal angle dependence of the form $\cos 2 \phi$ and $\sin 2\phi$
enters the differential cross section for fermion pair production in
$e^+e^-$ collisions, when the $s$-channel exchange of a tower of massive
gravitons is introduced. Here $\phi$ is the azimuthal angle of the
fermion defined relative to the $e^-$ beam axis as the $x$ axis, and the
$e^-$ transverse polarization direction as the $y$ axis. 
The coupling of gravitons at the leptonic
vertex has the form
\begin{equation}
\Gamma_{\mu\nu} = \gamma_\mu q_\nu + \gamma_\nu q_\mu,
\end{equation}
So far as the Dirac structure is concerned, there is only one $\gamma$
matrix in each term, and so this falls in the category of vector 
interactions. The four-vector $q_\nu$ appearing in the coupling could be
absorbed in the hadronic tensor. Thus, we could use the results of the
third and seventh rows of Table 3 to deduce the $\theta$ and $\phi$
dependence of the angular distribution for any final state.

In a recent work \cite{hep-ph/0511188}, the authors have considered
the possibility of studying non-commutative extension of the standard model
through $Z\gamma$ production at the Tevatron and the LHC.
It would be possible to extend this discussion to the
possibility of probing an underlying non-commutative field
theory at the linear collider with polarized beams.
They note that to leading order in the non-commutativity parameter,
contact interactions as well as triple-gauge couplings 
are introduced to describe
the process $f\bar f\to Z\gamma$. The technique developed here would be
useful to study angular distributions for the case of $e^+e^- \to Z
\gamma$ in this model.

We now apply an extension of our results to two examples where the
underlying theory is not the SM, but the 
MSSM. The first example is of neutralino pair production,
already alluded to in Sec. 4. In this case there is an $s$-channel as
well as $t$- and $u$-channel contributions. The $t$- and $u$-channel 
contributions can be
written using Fierz transformation in a form which has only $V$ and
$A$ couplings to the electron current \cite{Choi}, but with an overall
factor proportional to $(t-u)$ coming from the propagators. We can then 
apply our
results obtained for $V,A$ BSM interactions from Table 3 and 4. The case
1 considered in Sec. 4 for self-conjugate particles applies in this
case. We then find that it is possible to have a CP-odd correlation
given by a product of the correlation in the last line of either Table 3
or 4 and the factor $\vec p \cdot \vec K$, where $p$ is the momentum of
a neutralino. This has been discussed in ref. \cite{Choi,Bartl1} and
a numerical study of a corresponding CP-odd asymmetry has been carried
out in ref. \cite{Bartl1}.  

The other example from MSSM is that of chargino pair production.
This corresponds to the case 2b discussed in Sec. 4. We also refer
the reader to the comments in Sec. 5 regarding the C transformation
of the propagators entering the amplitude in this case.
We find
that for the process $e^+e^-\to \tilde\chi^+_i \tilde\chi^-_j$, where
$\tilde\chi^{\pm}_i$ denote charginos, no CP-odd correlation exists at
tree level \cite{Bartl2}. This can be understood from  our discussion
earlier where we found that in
the $V,A$ case, to which our present case can be reduced using a Fierz
transformation, the only CP-odd correlation arises from the last lines
in Tables 3 and 4, in the combination $\Delta\sigma_-$ corresponding to
the difference of the cross sections for $e^+e^-\to \tilde\chi^+_i
\tilde\chi^-_j$ and $e^+e^-\to \tilde\chi^+_j \tilde\chi^-_i$.
However, it is found in \cite{Bartl2} that the corresponding coefficient
vanishes at tree level.

Another example from the MSSM which would fall in the category of case 2b of
Sec. 4 is the neutralino pair-production process mentioned above, but
when the energy and momentum that is measured is of a lepton arising in
the decay chain of a neutralino. In this case, it is possible to
construct a CP-odd correlation using leptons of opposite charges, and
the effect is non-vanishing \cite{Bartl1}.

\section{Conclusions}
To conclude, we have considered the
exploration of the space-time structure of new physics beyond the
standard model in polarized $e^+ e^-$ annihilation at linear
collider energies.  We have studied the interference of the
standard model processes for a final state whose momentum alone
is measured and have considered the most general possible polarization
of the electron and positron beams.  Our work is the analog of
the exploration of the space-time structure of the neutral current
due to the $Z$ boson from its interference with the QED amplitude
for a given final state.  While our work is a logical extension of the
work of Dass and Ross,  it has several novel
features which were not present in that work.   We have provided a
significant extension not precedented in the literature in our discussion
of the case of
SM amplitudes which occur with fermion exchange 
in the $t$ and $u$ channels, by showing that they may be written in a
form
analogous to those with $s$-channel amplitudes, albeit with
momentum dependent form factors.  
We have also shown that some features of our treatment can be carried
over to an extension of SM, like MSSM, using as illustrations chargino
and neutralino pair production.
We have also considered
popular scenarios for BSM physics, resulting from either extra
dimensional models or from non-commutative models. 

\section{Acknowledgements} We thank the organizers and sponsors (BRNS) 
of WHEPP8 (Workshop on High Energy Physics Phenomenology 8) 
held at IIT Mumbai in January
2004, where this work was initiated.  
We thank Alfred Bartl for a careful reading of the manuscript and
useful suggestions.
BA thanks N.D. Hari Dass for a discussion.
BA also thanks the Council for Scientific and
Industrial Research for support during the course of these
investigations under scheme number 03(0994)/04/EMR-II, as well
as the Department of Science and Technology, Government of India.

\end{document}